\documentclass[12pt]{extarticle}
\usepackage[margin=36mm]{geometry}
\usepackage[english]{babel}
\usepackage[utf8]{inputenc}
\usepackage{amsmath}
\usepackage{graphicx}
\usepackage[colorinlistoftodos]{todonotes}
\usepackage{amsmath}
\usepackage{amssymb}
\usepackage{yfonts}
\usepackage{amsmath}
\usepackage{verbatim}
\usepackage[T1]{fontenc}
\usepackage[utf8]{inputenc}
\usepackage{authblk}

\providecommand{\keywords}[1]
{
  \small	
  \textbf{\textit{Keywords---}} #1
}
 
\title{Experimental parameters' Uncertainty limits for z-scan and f-scan techniques}

\author[1]{Esteban Marulanda\thanks{esteban.maruladaa@udea.edu.co}}
\author[1]{Edgar Rueda
\thanks{edgar.rueda@udea.edu.co}}
\affil[1]{Grupo de Óptica y Fotónica, Instituto de Física, Universidad de Antioquia U de A, Calle 70 No. 52-21, Medellín, Colombia}

\date{\today}

\begin{document}
\maketitle
\begin{abstract}
 In this paper, we present an analytical study of the relationship between the statistical distribution of a physical parameter and the uncertainties in the physical quantities used to determine it through indirect measurement. We investigate two possible methods for determining the physical quantity: linear regression and inversion of the equation in the parameter. Our analysis focuses on finding the limits of "small" uncertainties to guarantee a Gaussian distribution to the indirect physical quantity. Also, we introduce the "reliability cone" concept to describe the dependence of errors on the physical parameters uncertainties. We propose a new probability distribution for significant uncertainties and define the first three moments. We apply these methods to the z-scan and f-scan techniques, presenting the most sensitive parameters for the nonlinear two-photon absorption coefficient measurement. Finally, we implement our findings on experimental data of the two-photon absorption coefficient in CdSe.
\end{abstract}

\keywords{Two-photon absorption, z-scan, f-scan, uncertainty analysis, statistical distribution}

\section{Introduction}
\label{sec:intro}

When the expected value of a physical quantity is reported, it has only meaning if it is accompanied by its uncertainty \cite{Squires2001}. This is because there is no absolute certainty that the expected value coincides with the actual value of the physical quantity. Thus, it is only known with a certain probability that the actual value is within the interval defined by the uncertainty. Assigning a probability requires knowing the statistical distribution of the physical quantity. For quantities measured directly, even if their combined uncertainty is obtained from uncertainties with different statistical distributions, if certain conditions are followed, the guide to the expression of uncertainty in measurements (GUM) states that the distribution of the quantity can be approximated to a normal distribution  \cite{GUM2008}. Nevertheless, most of the physical quantities of interest are not obtained by direct measurements. Instead, they are derived indirectly from relations between other quantities. The problem is that the statistical distributions of quantities derived indirectly are normally distributed only in cases where the direct quantities from which it is derived have relative uncertainties much smaller than unity \cite{Silverman2004}.

Two examples of physical quantities measured indirectly can be found in the z-scan and f-scan techniques when measuring the nonlinear optical properties of materials. Z-scan is a technique of extended use thanks to its experimental simplicity compared to other techniques. It has been extensively used to measure the nonlinear properties in semiconductors, dielectrics, crystals, organic molecules, perovskites, and graphene, among others \cite{Sheik1990,Chapple1997,Dinu2003}. In principle, to obtain the nonlinearity, changes in the sample transmittance are measured while a laser beam irradiance varies. The sample is shifted through the optical axis of a convergent lens to vary the laser irradiance over the sample; thus, the technique's name. Knowing the transmittance as a function of the sample position $z$ it is possible to derive the nonlinear property with a theoretical model. F-scan is a modification to z-scan where a variation of the convergent lens focal distance replaces the displacement of the sample. In this architecture, the sample stays in a fixed position \cite{Rueda2019,Kolkowski2014}. A tunable lens is used to achieve the change in the lens focal length. Also, to consider the change in the focal distance and the sample fixed position, the theoretical model of z-scan is modified. Both techniques require many experimental parameters like the laser pulse temporal width or the sample thickness to obtain the nonlinearity.

Despite the relative simplicity of z-scan, much care must be taken in measuring the transmittance and the experimental parameters. A careless measurement will not attain a reliable nonlinear coefficient value \cite{Chapple1997,Aloukos2003,Rumi2010,Slavinskis2011}. Several works in the literature have studied the influence of the incorrect estimation of the experimental parameters in the correct estimation of the nonlinear property. The most studied parameters are the laser pulse temporal profile \cite{Chapple1997}, the laser pulse spatial profile \cite{Chapple1997,Rumi2010,Slavinskis2011}, the laser pulse repetition rate \cite{Chapple1997}, and the pulse temporal width \cite{Aloukos2003}. F-scan is a relatively new technique; thus, it has been reported in a limited number of articles \cite{babu2020third,arendt2015spectral,rueda2022nonlinear,koirala2020bimetallic}. In reference \cite{Rueda2019}, a proposal to include the uncertainties in the experimental parameters is presented, while in reference \cite{henao2022influencia}, the influence of each parameter and their correlations to the uncertainty of the nonlinear two-photon absorption is studied.

In this work, we study the influence of each experimental parameter on the precision and statistical distribution of the measured nonlinear two-photon absorption (TPA). We define the maximum admissible uncertainties that ensure a Gaussian distribution for the measured TPA. We also derive the correct statistical distribution for experiments with uncertainties greater than the limit imposed by the Gaussian distribution criteria. This analysis is essential for researchers reporting nonlinear properties measured with z-scan and f-scan.

The study is done for two different approaches in the data processing step toward determining the TPA. The first approach uses an inverse function \cite{Garcia2020}, and the second uses a curve-fitting linear regression \cite{Sheik1990,Rueda2019}.

In section two, a general analytical approach to the definitions of the distribution and uncertainty of the indirect physical quantity is derived for the particular case when the uncertainties of the direct variables are significantly small. Section three presents the z-scan and f-scan techniques. Sections four and five present the conditions need it to guarantee a Gaussian distribution for the indirect variable in z-scan and f-scan, respectively. Section sixth expands the analytical approach for the case where the conditions on the direct variables' uncertainties are not fulfilled. Finally, section seventh presents an example with reported CdSe experimental data.


\section{Methods}

Consider a physical experiment that is subject to governing physical law. In many cases, the relationship between the dependent variable $y$ and the independent variable $x$ can be described using differential operators represented by the equation $E(y,x,D)=0$, which characterizes the experiment \cite{Lombardi2016TheRO}. If an analytic solution or approximation can be obtained for this equation, we can then express the experiment using a corresponding mathematical equation,

\begin{equation}
\label{m1}
y=g(x,\beta;\{v_i\}_{i=1}^{M}),
\end{equation}

\noindent In this experiment, we aim to determine a physical parameter $\beta$ using a set of $M$ independent parameters $\{v_i\}_{i=1}^{M}$, which were measured by another experimental method. Equation \eqref{m1} relates these parameters to $\beta$ and is commonly used in experiments with an indirect measurement objective. Due to the stochastic nature of measurements in any experimental situation, the variables and quantities related to equation \eqref{m1} are random. Each variable has a particular probability density function (pdf) and a set of population parameters $\theta$ that characterize the distributions. Therefore, it is crucial to understand the dependence of the uncertainty of the distributions associated with $x$, $y$, and ${v_i}$ on the reliability of the value we report as our measurement of $\beta$.
The estimated value and error of $\beta$ will depend on how it is calculated. To address this issue, we consider two approaches: inversion of equation \eqref{m1} and a regression method. The following discussion will refer to random variables using capital letters.

\subsection{Method 1: inverse function}
 In the case where equation \eqref{m1} can be inverted in $\beta$, the physical parameter is determined according to

\begin{equation}
\label{m2}
\beta=g^{-1}(x,y;\{v_i\})\equiv f(x,y;\{v_i\}) .
\end{equation}

In its random form, it is written as follows:

\begin{equation}
\label{m3}
B= f(X,Y;\{V_i\}) ,
\end{equation}

\noindent where for each random variable of the experiment, we have $X\sim p_{X}(\theta_X)$, $Y\sim p_{Y}(\theta_Y)$, and $V_i\sim p_{V_i}(\theta_{V_i})$. If these parameters were determined using a direct measurement or can be approximated to a normal distribution, we have $\theta_X=(\mu_X,\sigma^{2}_X)$, $\theta_Y=(\mu_Y,\sigma^{2}_Y)$ and $\theta_{V_i}=(\mu_{V_i},\sigma^{2}_{V_i})$. In the particular case where equation \eqref{m3} has $M$ $V_i$ parameters and can be approximated to first order in a Taylor expansion,

\begin{equation}
\label{m4}
B\approx f(\mu_X,\mu_Y;\{\mu_{V_i}\})+\frac{\partial \beta}{\partial x}(X-\mu_{X})+\frac{\partial \beta}{\partial y}(Y-\mu_{Y})+\sum_{i=1}^{M}\frac{\partial \beta }{\partial V_i}(V_i-\mu_{V_i}) ,
\end{equation}

\noindent $B$ preserves the form of the distribution \cite{Taylor1997-ej}: $B\sim  N(\beta_{est},\sigma_{est}^{2})$, with the expected value

\begin{equation}
\label{ec-valM1}
\beta_{est}=f(\mu_x,\mu_y;\{\mu_{v_i}\}) ,
\end{equation}

\noindent and uncertainty equal to

\begin{equation}
\label{ec-desvM1}
 \sigma_{est}^{2}=\left(\frac{\partial \beta}{\partial x} \sigma_{X}\right)^{2}+\left(\frac{\partial \beta}{\partial y} \sigma_{Y}\right)^{2}+\sum_{i=1}^{N}\left(\frac{\partial \beta}{\partial V_i} \sigma_{V_i}\right)^{2} .
\end{equation}
 
Now, because we know the distribution, we can give a confidence interval in terms of the uncertainty of the parameters that will define the acceptability of the physical quantity: 

\begin{equation}
\label{m6}
\left |\beta_{est} -\beta_{real}  \right | \leq  t\sigma_{est},
\end{equation}

\noindent where $t$ is some factor, usually equal to one. This equation denotes what we call the acceptability cone: if the parameters $\beta_{est}$ and $\sigma_{est}$ are measured such that we have a measurement inside the cone, we can have certainty that, statistically speaking,  we can accept our measurement. Therefore, we need 2(M+1) population parameters to specify a point inside the cone.

\subsection{Method 2: linear regression.}
Suppose now we use a regression method for estimating the value of the physical parameter. For simplicity, suppose that equation \eqref{m1} is such that $g(x,\beta;\{v_i\})=\beta r(x;\{v_i\})+h(x;\{v_i\})$.  Let $n$ be the number of independent measurements labeled by $y_l$ and $x_l$. Because our purpose is to compare both methods in equal conditions, we use the principle of least squares \cite{Bain1991-mv} explicitly in order to give an analytic expression for computing the distribution of the physical parameter, instead of using an algorithm, that is

\begin{equation}
\label{m8}
S=\sum_{l=1}^{n} \left[ y_l-g(x_l,\beta; \{ v_{i}\}) \right]^{2}= \sum_{l=1}^{n} \left[y_l-\beta r(x_l; \{ v_{i}\})- h(x_l;\{v_i\})\right]^{2} .
\end{equation}

To estimate $\beta$ we minimize equation \eqref{m8} and obtain, in its random variable form,

\begin{equation}
\label{m9}
B=\frac{\sum_{l=1}^{n}r(X_l;\{ V_{i}\}) \left[Y_l-h(X_l;\{ V_{i}\})\right]}{\sum_{l=1}^{n}r^{2}(X_l;\{ V_{i}\})}.
\end{equation}

This equation as in the previous perspective defines a probability distribution that depends on the population parameters $\{ \theta_{X_l}\}_{l=1}^{n}$, $\{ \theta_{Y_l}\}_{l=1}^{n}$ and $\{ \theta_{V_i}\}_{i=1}^{M}$. Then, the specification of a point inside the acceptability cone depends on more parameters than in the previous case. To see this, consider the situations of small errors and moreover $X_l\sim N(\mu_{X_l},\sigma_{X_l}^{2})$, $Y_l\sim N(\mu_{Y_l},\sigma_{Y_l}^{2})$, $V_i\sim N(\mu_{V_i},\sigma_{V_i}^{2})$, that means, expanding equation \eqref{m9} in a similar way as equation \eqref{m4} , we get $B\sim  N(\beta_{est} \left(\{\mu_{x_l}\},\{\mu_{y_l}\};\{\mu_{v_i}\}\right),\sigma^{2}_{est} )$ with 

\begin{equation}
\label{ec-desvM2}
\sigma_{est}^{2}=\sum_{l=1}^{n}\left(\frac{\partial \beta}{\partial x_l} \sigma_{X_l}\right)^{2}+\sum_{l=1}^{n}\left(\frac{\partial \beta}{\partial y_l} \sigma_{Y_l}\right)^{2}+\sum_{i=1}^{M}\left(\frac{\partial \beta}{\partial v_i} \sigma_{V_i}\right)^{2} .
\end{equation}

Therefore, we need 2(M+2n) parameters to specify a point inside the cone.

\section{Z-scan and F-scan techniques}
\begin{figure}[ht]
\centering\includegraphics[width=14 cm]{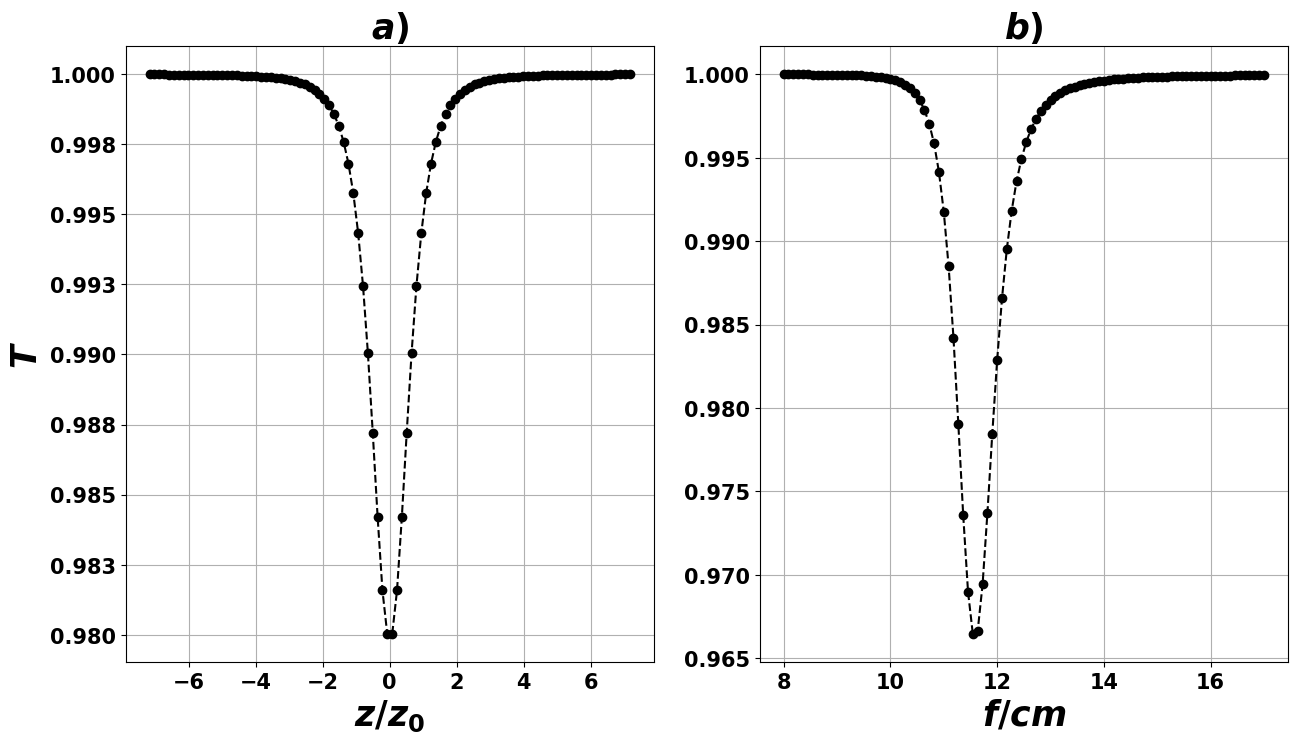}
\vspace{0.5cm}
\caption{(a) Simulated z-scan and (b) f-can signals for the parameters listed in table \ref{tabla-31}.}
\label{31}
\end{figure}

In z-scan, changes in the sample transmittance $T$ are related to the position $z$ of the sample and the rest of the optical parameters through the relation \cite{Dinu2003}:

\begin{equation}
\label{e31}
T=1-\frac{1}{2\sqrt{2}}\frac{\beta I_0(1-R)L_{eff}}{1+\left(\frac{z}{z_0} \right)^{2}} ,
\end{equation}

\noindent where $\beta$ is the TPA coefficient, $R$ is the normal reflectance of the sample, $L_{eff} = \big(1 - \exp{(-\alpha L)}\big)/\alpha$, $L$ is the sample thickness, $\alpha$ is the sample linear absorption, $I_0 = 4\sqrt{\ln{2}} P/ \big( \tau \nu \pi^{3/2} w^2(z) \big)$ is the optical-axis laser irradiance on the sample, $P$ is the average laser power, $\tau$ is the FWHM of the Gaussian temporal profile, $\nu$ is the laser repetition rate, and $w$ is the beam waist given by the relation

\begin{equation}
\label{ec-W}
w(z) = w_0 \sqrt{1+\Bigg(\frac{z}{z_0}\Bigg)^2} ,
\end{equation}

\noindent where $z_0 = \pi w_0^2/\lambda$ is the Rayleigh range, $w_0 = 2\lambda f C_f / \pi D$ is the Gaussian beam waist, $f$ is the tunable lens focal distance, $D$ is the beam diameter at the tunable lens and $C_f$ is a correction factor related to the Gaussian beam propagation factor \cite{Rueda2019}. In Figure \ref{31}a a typical z-scan signal is presented.

\begin{table}[h]
\centering
\begin{tabular}{ |c|c|c||c|c|c| } 
 \hline
 Parameter & Unit & Value & Parameter & Unit & Value \\ 
 \hline
 $\beta$ & cm/GW & 3.4 & $L$ & mm & 1.0 \\ 
 $P$ & mW & 200 & $\tau$ & fs & 200 \\ 
 $\lambda$ & nm & 790 & $D$ & mm & 2.0 \\
 $f$ & cm & 15 & $C_f$ &  & 1.36  \\
 $R$ & & 0.33 & $\nu$ & MHz & 90.6 \\
 $\alpha$ & 1/m & 0.26 & $d$ & cm & 11.6 \\
 \hline
\end{tabular}
\caption{Experimental parameters for simulated z-scan and f-scan signals.}
\label{tabla-31}
\end{table}

For the case of f-scan, equation \eqref{e31} is used by replacing the axial position $z$ for $d_s - f$, where $d_s$ is the distance between the sample and the tunable-focus lens and $f$ is the tuned focal distance: 

\begin{equation}
\label{ec-fscan}
T=1-\frac{1}{2\sqrt{2}}\frac{\beta I_0\big(1-R\big)L_{eff}}{1+\left(\frac{d_s-f}{z_0(f)} \right)^{2}} .
\end{equation}

Figure \ref{31}b shows a typical signal. Notice that the f-scan signal is asymmetric because the focal distance, and thus the beam waist and Rayleigh range, constantly changes.

\section{Results for z-scan}
In this section, the simulated signal $T$ of figure \ref{31}a will be considered the real signal. The simulated experimental signal is generated by introducing a random Gaussian noise to the transmittance of the real signal. Next, to measure $\beta$, the experimental signal and experimental parameters are used, but the experimental parameters are supposed not known with absolute certainty; they are considered to have a random Gaussian uncertainty. Because the $T$ is noisy and there is uncertainty on the value of the parameters, it is expected that $\beta$ will be obtained with a relative error with respect to the real value. Thus, the expected value of $\beta$ is reported with uncertainty.

\subsection{Method 1: inverse function}
To use method 1, from equation \eqref{e31}, $\beta$ is expressed in terms of the transmittance and the experimental parameters, 

\begin{equation}
\label{e32}
\beta=\left(\frac{1}{2\sqrt{2}} \frac{(1-R) I_0L_{eff}}{1+\left(\frac{z}{z_0}\right)^{2}} \right)^{-1} (1-T) .
\end{equation}

After measuring the transmittance for one $z$ position, the value of $\beta$ is obtained by substituting the expected values of the transmittance and the parameters in equation \eqref{e32}.

\begin{figure}[ht]
\centering\includegraphics[width=10 cm]{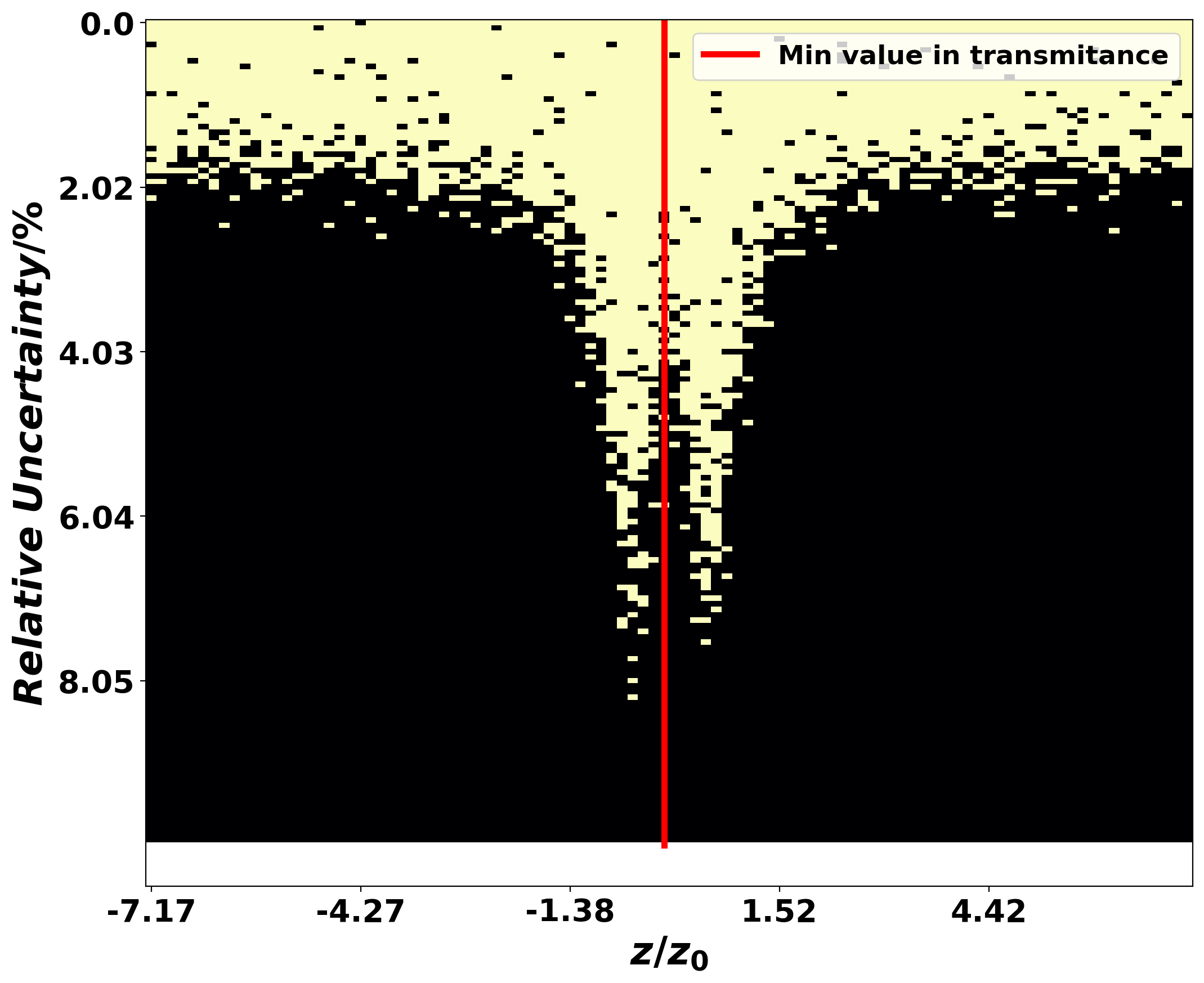}
\vspace{0.3cm}
\caption{Z-scan: conditions to secure a Gaussian distribution for $\beta$ as a function of the sample position $z/z_0$ and the "relative uncertainty" of the experimental parameters. Dark dots correspond to conditions where the hypothesis of a Gaussian distribution is rejected.}
\label{32}
\end{figure}

Knowing that the transmittance and all the experimental parameters are normally distributed and have the same relative uncertainty, a normal test is performed considering the null hypothesis that $\beta$ is normally distributed. The distribution of $\beta$ is constructed by randomly picking the values of the transmittance and parameters from their respective distributions. The result of the normal test is shown in figure \ref{32}. For all sample positions, the null hypothesis is not rejected for relative uncertainties under $2\,\%$. At minimum transmittance, the relative uncertainty can reach $4\,\%$, while it can go until $8\,\%$ for positions closed to the minimum transmittance.

\begin{figure}[ht]
\centering\includegraphics[width=13 cm]{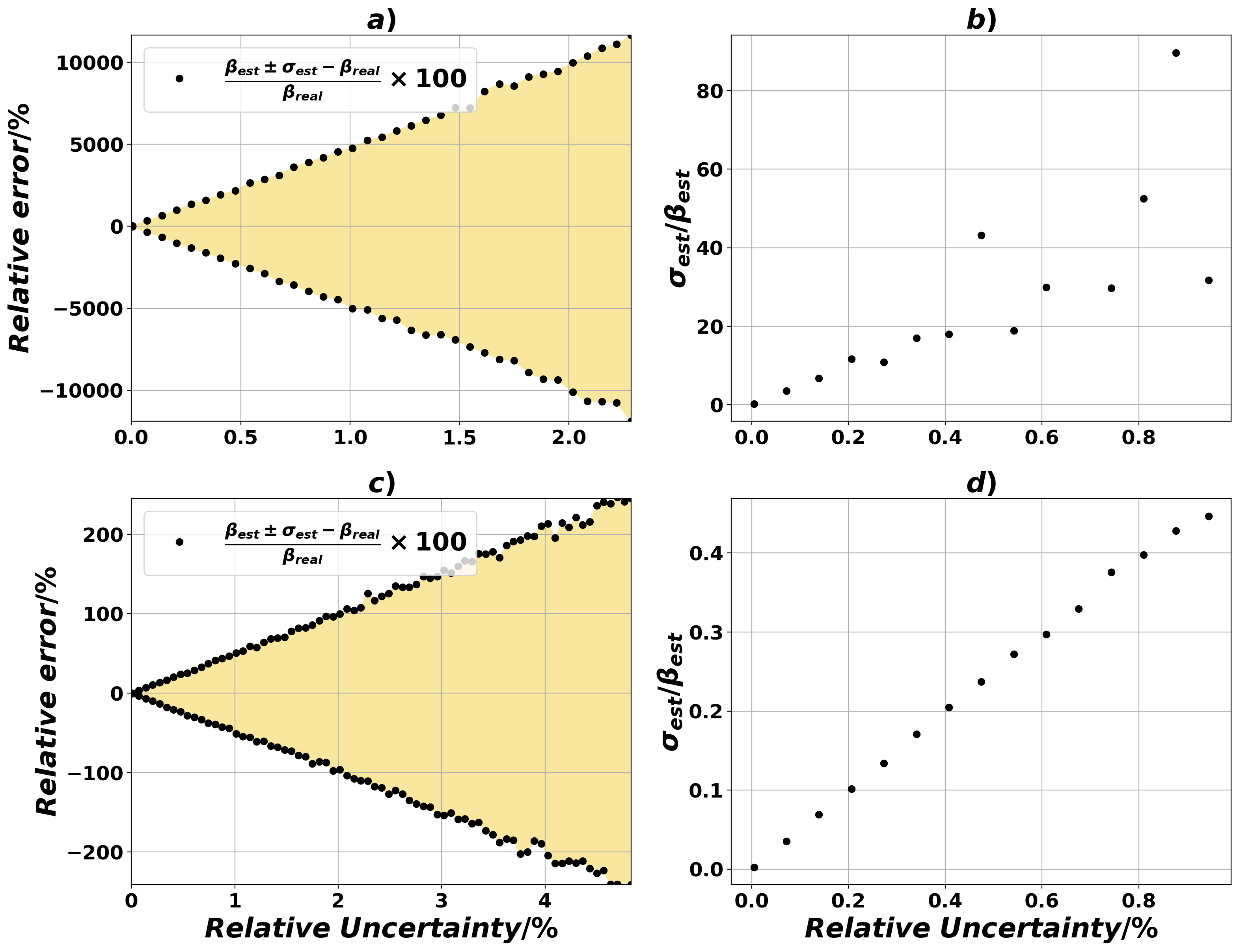}
\vspace{0.3cm}
\caption{Z-scan: $\beta$ relative error and relative uncertainty. All parameters and transmittance with the same relative uncertainty.  a) Relative error  and b) relative uncertainty at maximum transmittance $z/z_0=3$. c) Relative error and d) relative uncertainty at minimum transmittance $z/z_0=0$. }
\label{33}
\end{figure}

Figure \ref{33} shows the relative error and relative uncertainty of $\beta$. Here, the relative error corresponds to the difference to the real value in percentage:

\begin{equation}
\label{ec-relativeError}
Relative\, error = 100\times\frac{\beta - \beta_{real}}{\beta_{real}} .
\end{equation}

From figure \ref{33}, it is clear that for all positions, even for relative uncertainties of the experimental parameters smaller than $1\,\%$, the relative error is very high, making the measurement useless. Therefore, each parameter's influence is analyzed to understand why this is happening. 

\begin{figure}[ht]
\centering\includegraphics[width=14 cm]{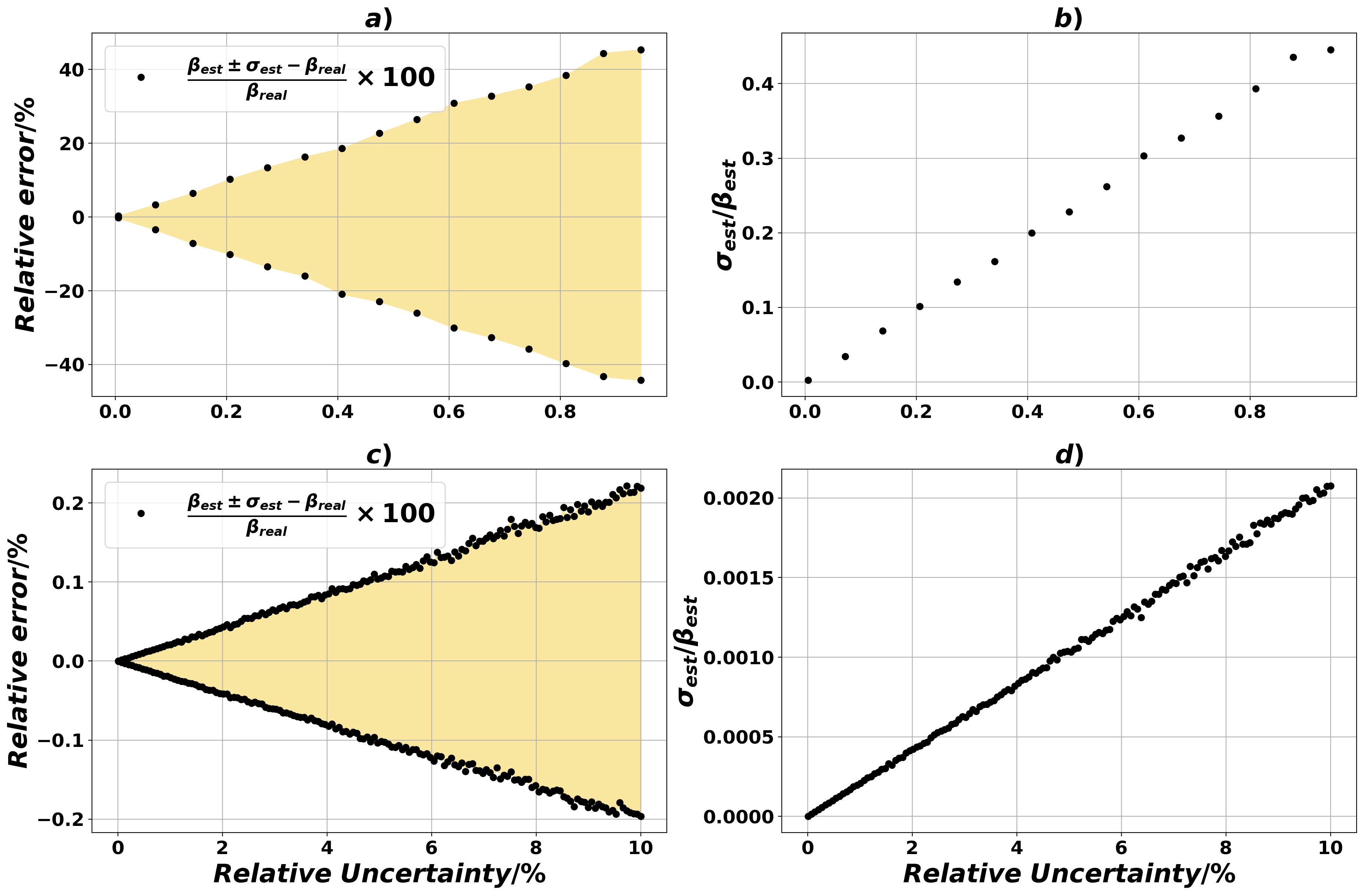}
\vspace{0.3cm}
\caption{Z-scan: $\beta$ relative error and relative uncertainty at minimum transmittance.  a) Relative error  and b) relative uncertainty for $T$ known with uncertainty. c) Relative error and d) relative uncertainty for $z$ known with uncertainty.}
\label{34}
\end{figure}

From figures \ref{34}a and b, it is clear that the transmittance value is responsible for the high inaccuracies and uncertainties in determining $\beta$. In contrast, for all the ranges where the distribution of $\beta$ is Gaussian, the rest of the parameters generate inaccuracies smaller than $20\,\%$ and relative uncertainties below $2\,\%$. As an example, figures \ref{34}c and d, show $\beta$ relative error and relative uncertainty when only the position $z$ has uncertainty.

\subsection{Method 2: linear regression}

\begin{figure}[ht]
\centering\includegraphics[width=13 cm]{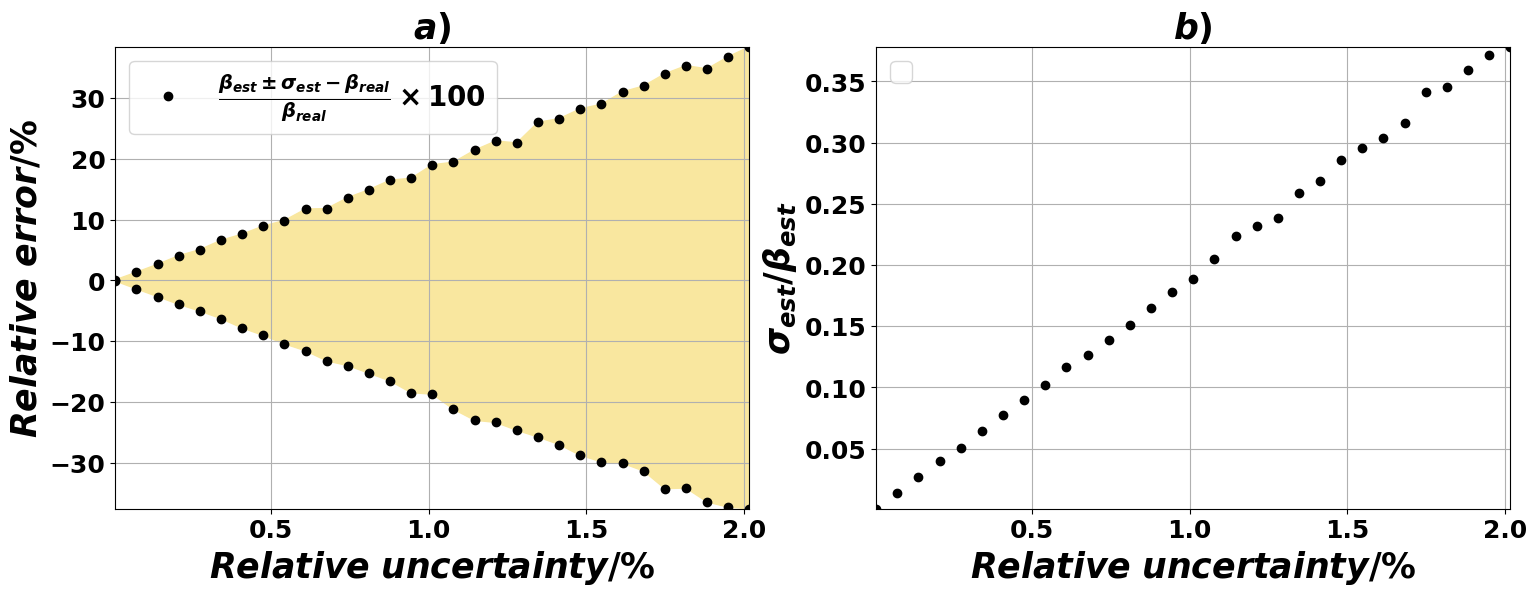}
\vspace{0.3cm}
\caption{Z-scan: relative error and relative uncertainty obtained with method 2. All parameters and the transmittance with the same relative uncertainty. $\beta$  a) relative error and b) relative uncertainty as a function of the relative uncertainty of the experimental parameters and the transmittance.}
\label{36}
\end{figure}

From equation \eqref{e31}, $r(x;\{v_i\})=-\frac{1}{2\sqrt{2}}\frac{ I_0(1-R)L_{eff}}{1+\left(\frac{z}{z_0} \right)^{2}}$ and $h(x,\{v_i\})=1$. Making the substitution in equation \eqref{m9} with the transmittance data and the experimental parameters and expected values, it is possible to obtain the best value for $\beta$. By repeating the procedure used for method 1, the distribution for $\beta$ is constructed by randomly picking values of the experimental parameters from their respective distributions. After performing the normal test, it is found that for relative uncertainties as high as $10\,\%$, the distribution of $\beta$ can be considered Gaussian. Nevertheless, like with method 1, $\beta$ relative errors and uncertainties are very high even for parameter's relative uncertainties smaller than $1\,\%$ (see figure \ref{36}).

\begin{figure}[ht]
\centering\includegraphics[width=13 cm]{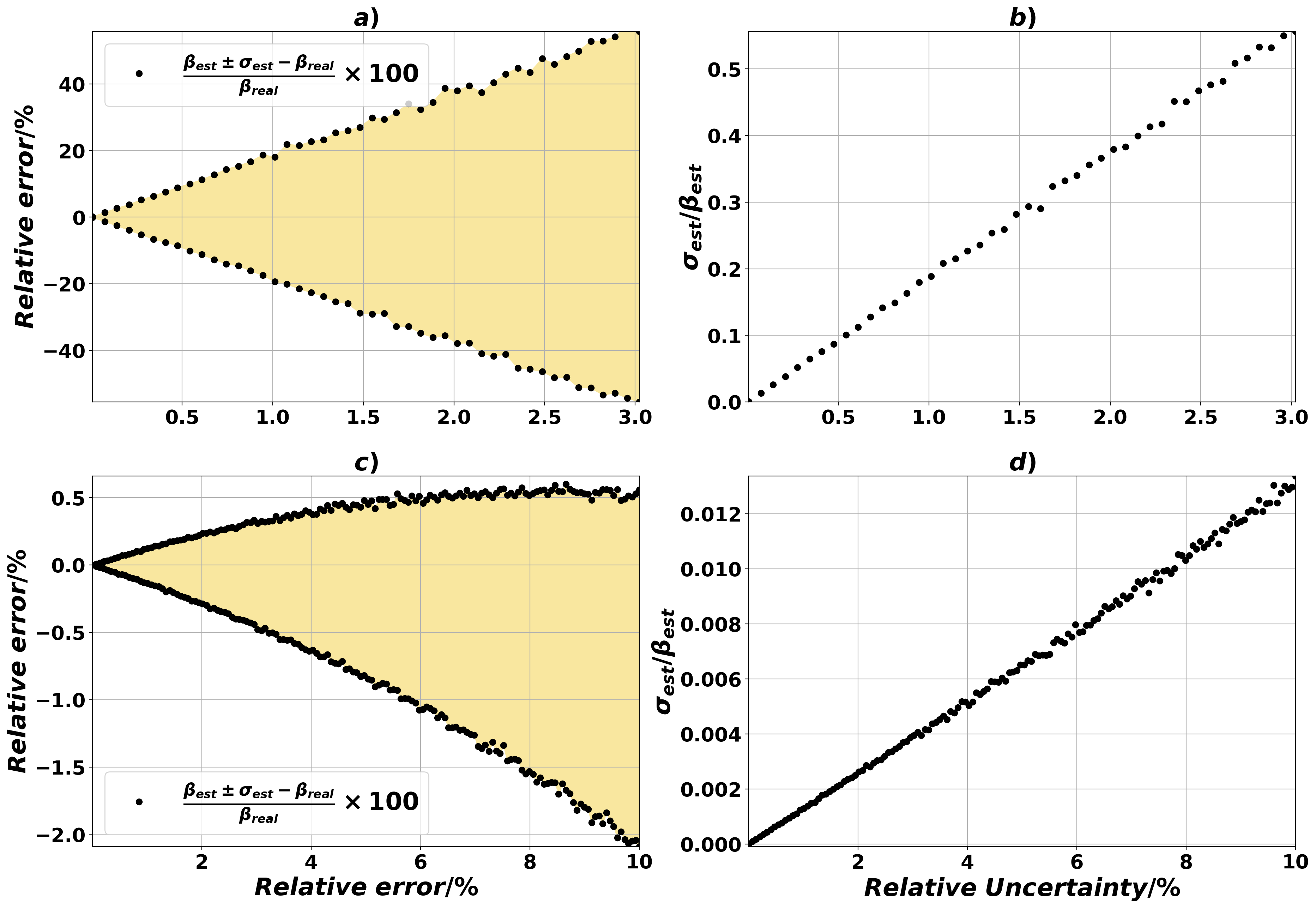}
\vspace{0.3cm}
\caption{Z-scan: $\beta$ relative error and relative uncertainty with method 2.  a) Relative error  and b) relative uncertainty; $T$ known with uncertainty. c) Relative error and d) relative uncertainty; $z$ known with uncertainty.}
\label{37}
\end{figure}

Again, the analysis is repeated, assigning uncertainty to only one variable; results are presented in figure \ref{37}. Similar to method 1, the relative error and uncertainty of $\beta$ are more prone to uncertainties in the transmittance. Nevertheless, this susceptibility to the transmittance uncertainty is not as high as in the case of method 1 because method 2 requires more transmittance values to obtain the value of $\beta$.

\begin{figure}[ht]
\centering\includegraphics[width=14 cm]{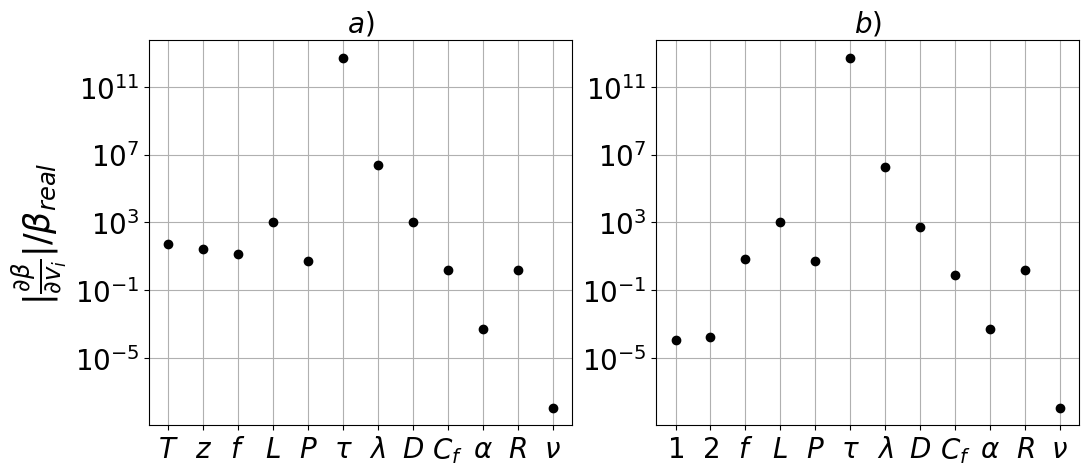}
\vspace{0.3cm}
\caption{Z-scan: adimensional derivatives of equations \eqref{ec-desvM1} and \eqref{ec-desvM2} for a) method 1 and b) method 2. In b), variables 1 and 2 correspond to the total contribution of all the transmittances $T_i$ and positions $z_i$, respectively.}
\label{compz}
\end{figure}

To compare the influence of the transmittance and experimental parameters uncertainties on the uncertainty of $\beta$, in figure \ref{compz} we have plotted the derivatives of equations \eqref{ec-desvM1} and \eqref{ec-desvM2}. To make the derivates adimensional, they are divided by $\beta_{real}$ and multiplied by the unit of their respective parameter uncertainty. From the figure, it is clear that the influence of the experimental parameters is the same for both methods. In contrast, the uncertainties of the transmittance and the sample position $z$ will be less critical in method 2 as the number of points of the linear regression is increased.

\section{Results for f-scan.}
In this section, the simulated signal $T$ of figure \ref{31}b will be considered the real signal. The simulated experimental signal is generated by introducing a random Gaussian noise to the transmittance of the real signal. Next, to measure $\beta$, the experimental signal and experimental parameters are used, but the experimental parameters are supposed not known with absolute certainty; they are considered to have a random Gaussian uncertainty. Because the $T$ is noisy and there is uncertainty on the value of the parameters, it is expected that $\beta$ will be obtained with a relative error with respect to the real value. Thus, the expected value of $\beta$ is reported with uncertainty.

\subsection{Method 1: inverse function}
\begin{figure}[ht]
\centering\includegraphics[width=13 cm]{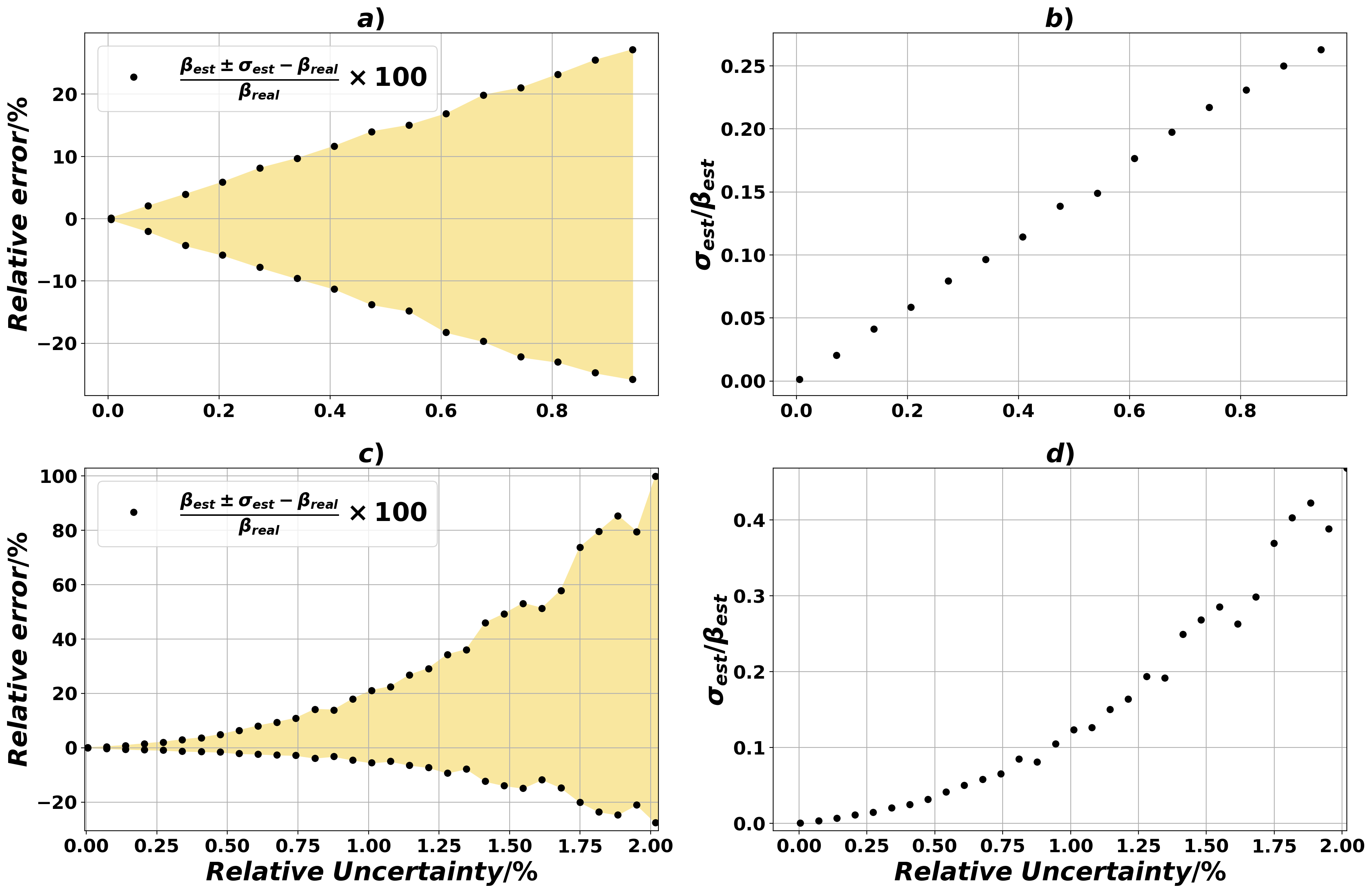}
\vspace{0.3cm}
\caption{F-scan: $\beta$ relative error and relative uncertainty at minimum transmittance.  a) Relative error  and b) relative uncertainty for $T$ known with uncertainty. c) Relative error and d) relative uncertainty for $f$ known with uncertainty.}
\label{42}
\end{figure}

To use method 1, from equation \eqref{42} $\beta$ is expressed in terms of the transmittance and the experimental parameters, 

\begin{equation}
\label{ec-42}
\beta=\left(\frac{1}{2\sqrt{2}} \frac{(1-R) I_0(f)L_{eff}}{1+\left(\frac{d_s-f}{z_0(f)}\right)^{2}} \right)^{-1}\left(1-T\right).
\end{equation}

After measuring the transmittance for one $f$ position, the value of $\beta$ is obtained by substituting the expected values of the transmittance and the parameters in equation \eqref{ec-42}.

\begin{figure}[ht]
\centering\includegraphics[width=12 cm]{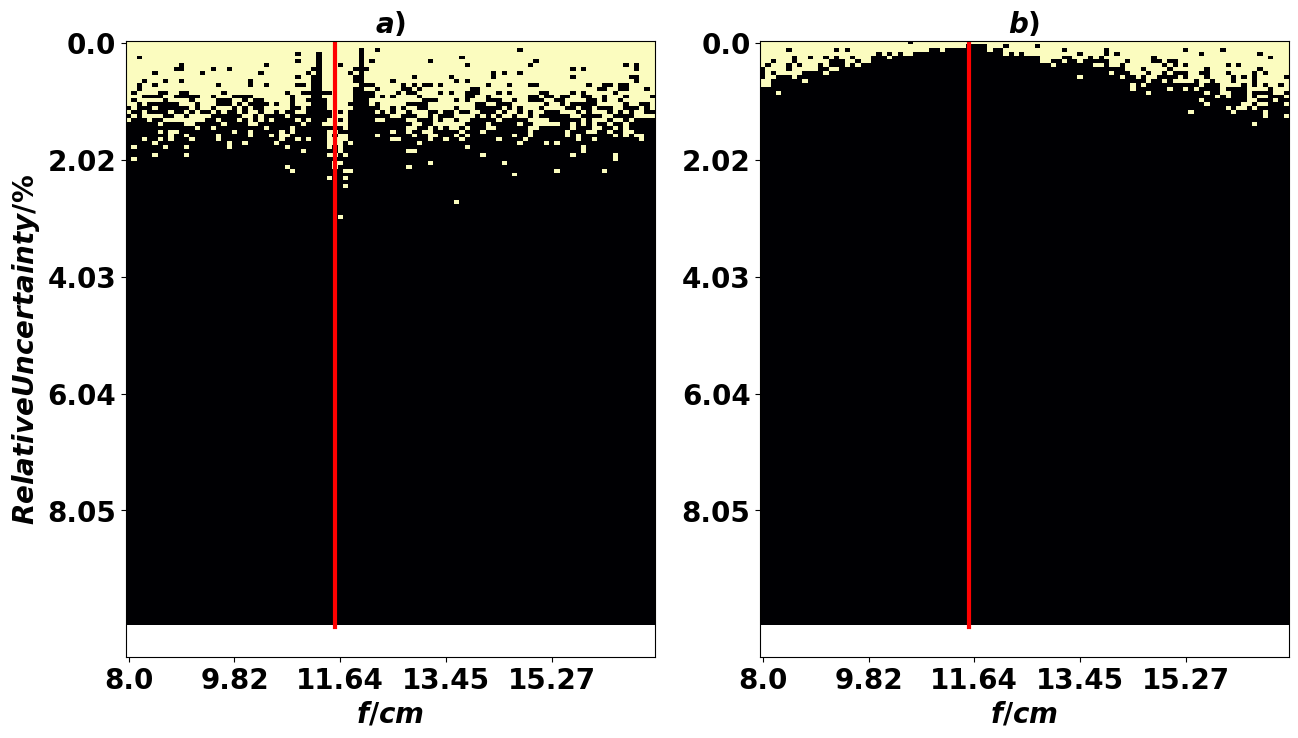}
\vspace{0.3cm}
\caption{F-scan: conditions to secure a Gaussian distribution for $\beta$ as a function of the tunable-lens focal distance $f$ and the "relative uncertainty" of the experimental parameters. Dark dots correspond to conditions where the hypothesis of a Gaussian distribution is rejected. Results when only a) $D$ or b) $d_s$ have uncertainties.}
\label{41}
\end{figure}

The transmittance and all the experimental parameters are normally distributed and have the same relative uncertainty. A normal test is performed considering the null hypothesis that $\beta$ is normally distributed. The distribution of $\beta$ is constructed by randomly picking up the values of the transmittance and parameters from their respective distributions. Similar to the case of z-scan, the Gaussianity of the distribution for $\beta$ is restricted to relative uncertainties in the parameters smaller than $1\,\%$, and smaller than $0.5\,\%$ when the measurements are done at the minimum transmittance. An analysis for each parameter indicates that the most critical experimental parameter is $d_s$, see figure \ref{41}b. Fortunately, the distance $d_s$ can be accurately directly measured from the minimum transmittance, where $d_s = f$. The rest of the parameters do not reject the null hypothesis for uncertainties under $2\,\%$ (see figure \ref{41}a for an example). 

Excluding from the analysis the distance $d_s$, in figure \ref{42}, the two most critical variables are shown: the transmittance $T$ and the tunable-lens focal distance $f$. It is possible to conclude that in order to secure a $\beta$ relative error under $20\,\%$ the transmittance has to be measured with a relative uncertainty smaller than $0.6\,\%$ and the focal distance with a relative uncertainty smaller than $1\,\%$.

\subsection{Method 2: linear regression}
\begin{figure}[ht]
\centering\includegraphics[width=13 cm]{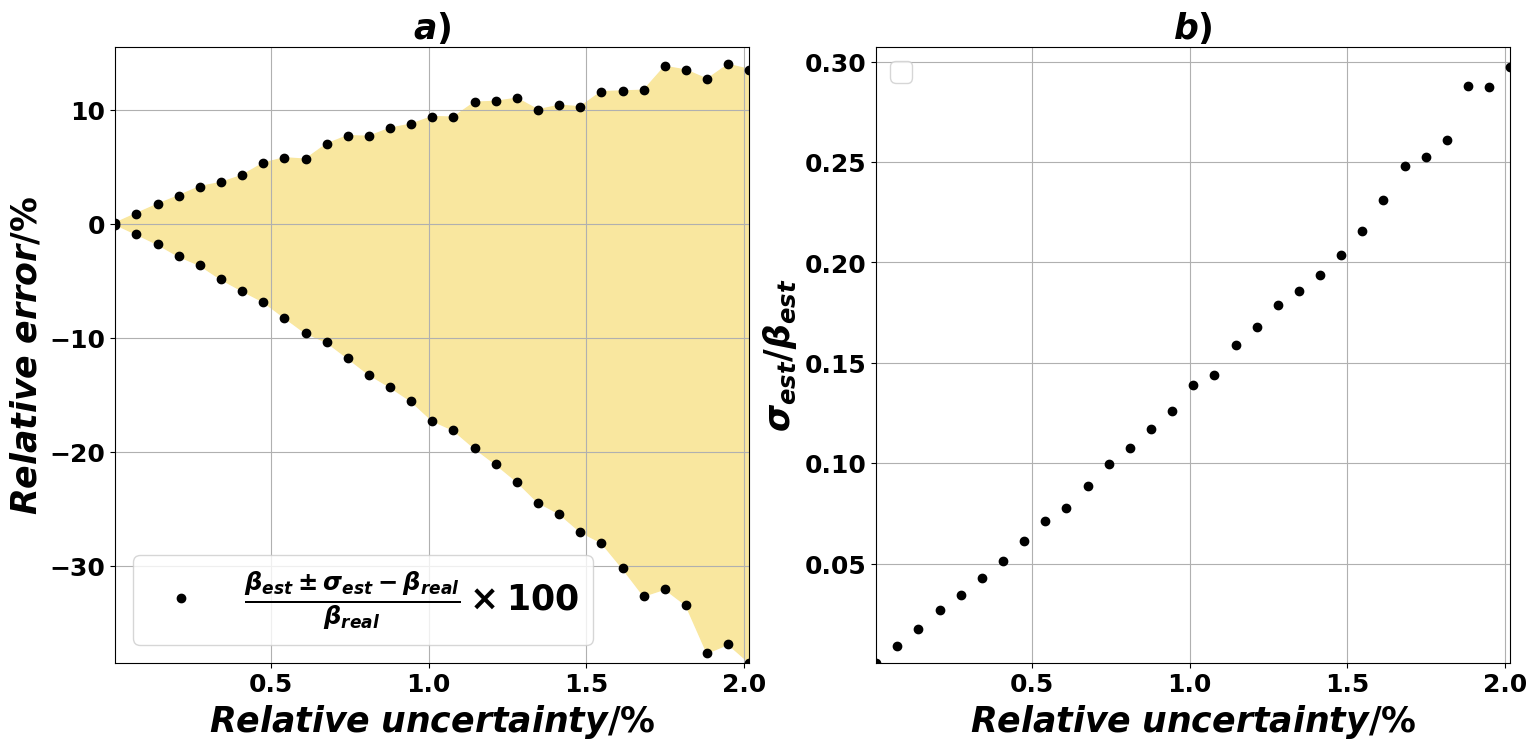}
\vspace{0.3cm}
\caption{F-scan: relative error and relative uncertainty obtained with method 2. All parameters and the transmittance with the same relative uncertainty. $\beta$  a) relative error and b) relative uncertainty as a function of the relative uncertainty of the experimental parameters and the transmittance.}
\label{43}
\end{figure}

From equation \eqref{ec-fscan}, $r(x;\{v_i\})=-\frac{1}{2\sqrt{2}}\frac{ I_0(1-R)L_{eff}}{1+\left(\frac{d_s - f}{z_0(f)} \right)^{2}}$ and $h(x,\{v_i\})=1$. Substituting in equation \eqref{m9} with the transmittance data and the experimental parameters' expected values, it is possible to obtain the best value for $\beta$. By repeating the procedure used for method 1, the distribution for $\beta$ is constructed by randomly picking up values of the experimental parameters from their respective distributions. After performing the normal test, it is found that for relative uncertainties as high as $10\,\%$, the distribution of $\beta$ can be considered Gaussian, having the same behavior encountered with z-scan. However, like with method 1, once more $\beta$ relative errors and uncertainties are of the order of $20\,\%$ even for the parameter's relative uncertainties of approximately $1\,\%$ (see figure \ref{43}).

\begin{figure}[ht]
\centering\includegraphics[width=14 cm]{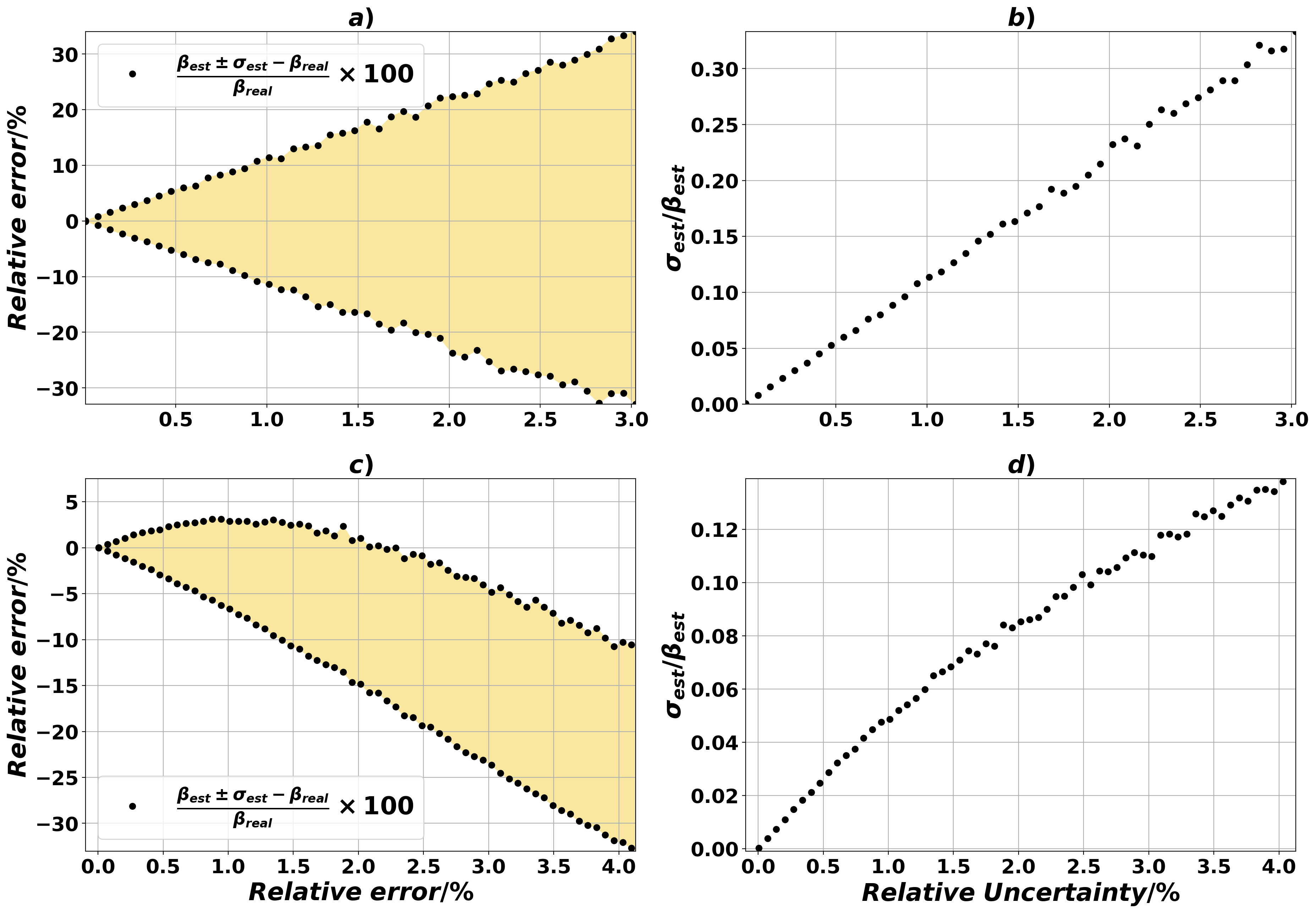}
\vspace{0.3cm}
\caption{F-scan: $\beta$ relative error and relative uncertainty with method 2.  a) Relative error  and b) relative uncertainty; $T$ known with uncertainty. c) Relative error and d) relative uncertainty; $f$ known with uncertainty.}
\label{fig-43}
\end{figure}

After analyzing each variable it is clear that all of them contribute similarly to the error and uncertainty of $\beta$; excluding the distance $d_s$. An example of two variables is shown in figure \ref{fig-43}.

Comparing the influence of all variables on the uncertainty of $\beta$ through the adimensional derivatives of equations \eqref{ec-desvM1} and \eqref{ec-desvM2}, the behavior is the same as for z-scan (see figure \ref{compz}). In this case, $f$ behaves exactly as $z$, and the value for $d_s$ is of the order of the value for $f$ in method 1.

\section{Distributions for not significantly small uncertainties}\label{sec-higher}

As we have shown, in many cases, we cannot express the reliability of the physical parameter in terms of an interval provided by a Gaussian distribution. Higher uncertainties indicate that the first-order approximation provided by equation \eqref{m4} is no longer accurate. In such cases, what can we do? For simplicity, we will consider a scenario where uncertainty in one parameter, say $v$, is only significant. In this case, we can use a second-order Taylor approximation to express an arbitrary function

\begin{equation}
\begin{split}
\label{m10}
B &\approx f(\mu_{v})+c_1 \left( {V}-\mu_{v}\right)+c_2(V-\mu_{v})^{2} \\ &=c+\left[\sqrt{c_2}\left(V-\mu_{v}\right)+\frac{c_1}{2 \sqrt{c_2}}\right]^{2},
\end{split}
\end{equation}

\noindent where $c=f(\mu_{v})-\frac{c_1^{2}}{4c_2}$. Let us rename $Y=\sqrt{c_2}\left(V-\mu_{v}\right)+\frac{c_1}{2 \sqrt{c_2}}$. This new random variable follows a Gaussian distribution, $Y \sim N(\mu_Y,\sigma^{2}_{Y})$, with $\mu_Y=\frac{c_1}{2 \sqrt{c_2}}$ and $\sigma^{2}_{Y}=c_2 \sigma_{v}$. Now, the cumulative distribution function of B,

\begin{equation}
\label{m11}
F_B(x)=P(c+Y^{2}<x)=P(\left | Y \right | < \sqrt{x-c}).
\end{equation}

Finally, if we take the derivative with respect to x, we find the pdf of the random variable B,

\begin{equation}
\label{m12}
\begin{split}
p_B(x) & =\frac{1}{2\sqrt{x-c}}\left( p_Y(\sqrt{x-c})+ p_Y(-\sqrt{x-c})\right)  \\ &
=\frac{1}{\sqrt{2 \pi \sigma^{2}_Y \left(x-c\right)}}\exp{\left( - \frac{1}{2 \sigma_Y^{2}}\left[ x-c+\mu_y^{2}\right]\right)}\cosh{\left( \frac{\mu_Y}{\sigma_Y^{2}}\sqrt{x-c}\right)}.
\end{split}
\end{equation}

To understand the difference between the corrected distribution and the Gaussian distribution, We calculate the moment-generating function

\begin{equation}
\label{m13}
\begin{split}
M_{B}(t)=\textbf{E}\left [ \text{exp}(tB) \right ]=\left(1-2 \sigma^{2}_{Y}t\right)^{-1/2}\exp{\left(\frac{t\mu^{2}_Y}{1-2 \sigma^{2}_{Y}t}+tc\right)}
\end{split}
\end{equation}

With this equation, we can compute the first three moments $m_i=\frac{\mathrm{d}^{i} M_B(t)}{\mathrm{d} t^{i}} |_{t=0}$ associated with the mean value, variance, and skewness:

\begin{equation}
\label{m14}
\begin{split}
m_1=\mu_B=f(\mu_{v})+c_2^{2}\sigma_v^{2},
\end{split}
\end{equation}

\begin{equation}
\label{m15}
\begin{split}
m_2=\sigma^{2}_B+\mu^{2}_{B}=c_1^{2}\sigma_v^{2}+3c_2^{4} \sigma_v^{4}+2c_{2}^{2}\sigma_v^{2}f(\mu_v)+f^{2}(\mu_v),
\end{split}
\end{equation}

\begin{equation}
\label{m16}
\begin{split}
m_3=15c_2^{6}\sigma_v^{6}+9c_2^{4}\sigma_v^{4}+f^{3}(\mu_v)+3c_2^{2}\sigma_v^{2}\left(c_1^{2}\sigma_v^{2}+f^{2}(\mu_v) \right )+3c_1^{2}\sigma_v^{2}\left(2+f(\mu_v) \right ).
\end{split}
\end{equation}

As is expected, the mean value and variance are corrected values of the Gaussian distribution. This correction depends on the parameter uncertainty and the square of the second derivative of the function that is used. Then, if $c_2^{2}\sigma_v^{2}\ll 1$ we can have certainty that the mean value and variance coincide with the Gaussian mean value. The fact that $m_3$ is in general different from zero implies that the pdf will not be symmetric \cite{Bain1991-mv}. In contrast two the first two moments, this will be true even for a small uncertainty in the parameter, because there is a dependence with $f^{3}(\mu_v)$.

\section{Application: CdSe two-photon absorption}
In order to see the effect of the distribution on the confidence interval of the uncertainty reported for the measured physical quantity, we will apply method 1 to the CdSe data reported in reference \cite{Rueda2019}. In table \ref{tabla-CdSe} we list the transmittance and experimental parameters taken from reference \cite{Rueda2019}.

\begin{table}[h!]
\centering
\begin{tabular}{ |c|c|c||c|c|c| } 
 \hline
 Parameter & Unit & Value & Parameter & Unit & Value \\ 
 \hline
 $T$ &  & 0.93 & $L$ & mm & $0.79 \pm 0.01$ \\ 
 $P$ & mW & $145 \pm 5$ & $\tau$ & fs & $71.0 \pm 0.3$ \\ 
 $\lambda$ & nm & $790 \pm 1$ & $D$ & mm & $2.0 \pm 0.8$ \\
 $f$ & mm & $109.0 \pm 0.5$ & $C_f$ &  & $1.36 \pm 0.04$  \\
 $R$ & & $0.185 \pm 0.005$ & $\nu$ & MHz & 90.9 \\
 $\alpha$ & 1/m & 3$69 \pm 37$ & $d_s$ & mm & $109.0 \pm 0.5$ \\
 \hline
\end{tabular}
\caption{CdSe experimental parameters for f-scan. Taken from reference \cite{Rueda2019}.}
\label{tabla-CdSe}
\end{table}

\begin{figure}[ht]
\centering\includegraphics[width=14 cm]{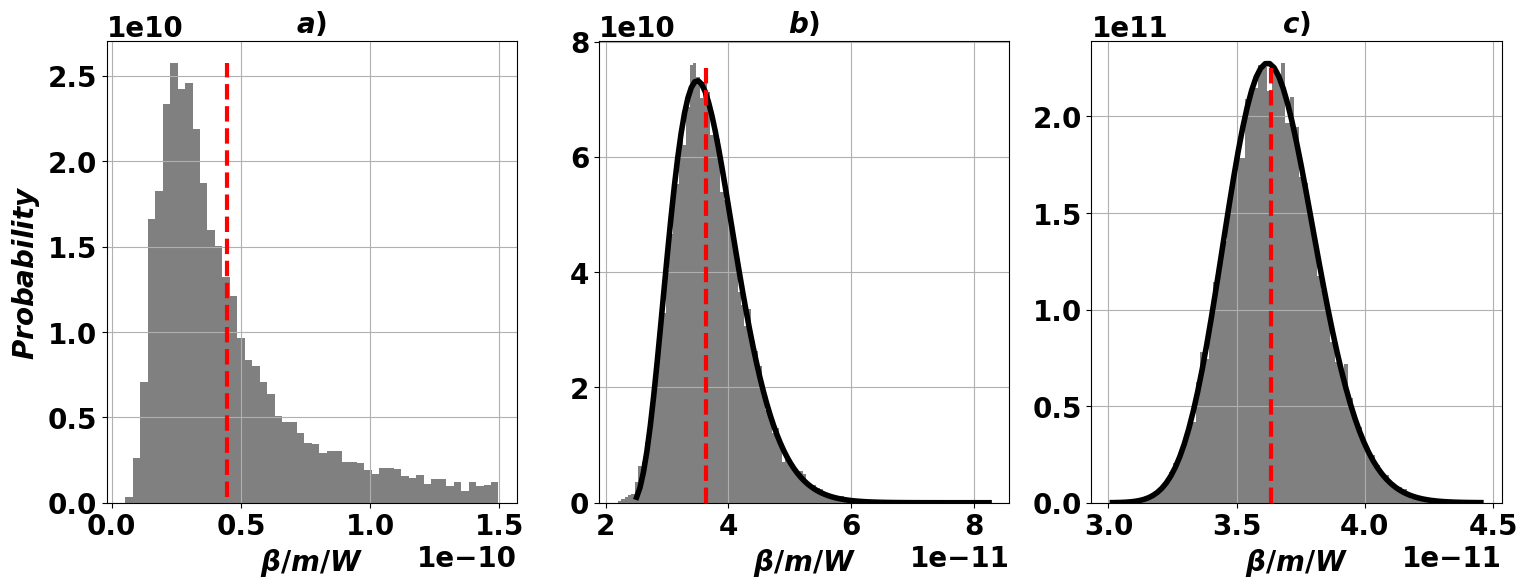}
\vspace{0.3cm}
\caption{Resulting distribution for CdSe experimental data. The black curve corresponds to equation \eqref{m12} and the red dashed line is the expected value of the distribution. In a) the uncertainties of all parameters are considered, in b) only a relative uncertainty of $8\,\%$ is considered for $D$, and in c) only a relative uncertainty of $2\,\%$ is considered for $D$.}
\label{fig-CdSe}
\end{figure}

Considering that all direct variables follow a Gaussian distribution, we artificially built the distribution of $\beta$ using the inverse function. In figure  \ref{fig-CdSe} the distribution using table \ref{tabla-CdSe} is plotted. It is clear that the distribution is not Gaussian, and because the relative uncertainty of $D$ is $40\,\%$ not even the approximation of section \ref{sec-higher} fits the distribution. The distributions of figures \ref{fig-CdSe}b and c were obtained considering only relative uncertainties of $8\,\%$ and $2\,\%$ for $D$, respectively.

In table \ref{tabla-resultados} we present the results for $\beta$ with the correct expected value, uncertainty, and probability of the confidence interval. For $\frac{\sigma_D}{D} = 40\,\%$ the relative uncertainty of $50\,\%$ for $\beta$ is too high; thus is not interesting. For the case of $\frac{\sigma_D}{D} = 8\,\%$, where $\beta = (3.6 \pm 0.6)\text{cm/GW}$, taking the value reported in reference \cite{Rueda2019} for f-scan of $(1.8 \pm 0.3)\text{cm/GW}$ the values disagree. But, surprisingly, it agrees with the value obtained with the differential f-scan technique of $(4.6 \pm 0.6)\text{cm/GW}$. 

\begin{table}[h]
\centering
\begin{tabular}{ |c|c|c|c| } 
 \hline
 $\sigma_D / D$ & $\beta_{est}$/cm/GW & $\sigma_{est}$/cm/GW & Certainty/$\%$ \\ 
 \hline
 $0.40$ & 4 & 2 & 74.1 \\ 
 $0.08$ & 3.6 & 0.6 & 64.4 \\ 
 $0.02$ & 3.63 & 0.18 & 64.6 \\
 \hline
\end{tabular}
\caption{$\beta$ results considering different relatives uncertainties for $D$. }
\label{tabla-resultados}
\end{table}

\pagebreak

\section*{Conclusions}
In this work, we developed a general analytical approach to defining the statistical distribution, expected value, and uncertainty of an indirect physical quantity. The derivation is made for two methods used to obtain the indirect quantity: an inverse function and a linear regression. First, we present the case where the uncertainty of the direct variables, from which the indirect variable is derived, is small enough to consider for the indirect variable a Gaussian distribution. Next, we find the expressions for the statistical distribution and the first three moments of the indirect variable when one of the direct variables does not satisfy the condition of small uncertainty. As expected, the expressions show that the expected variable is shifted concerning the variable predicted by a Gaussian distribution. 

The conditions to guarantee a Gaussian distribution, acceptable errors, and acceptable uncertainties in the indirect variable are studied for the z-scan and f-scan optical techniques. For both techniques, the transmittance is the most restricting variable when method 1 is used because, in method two, the restrictions over it are reduced due to the several measurements of transmittance that are needed. Nevertheless, when the transmittance is measured with significant precision, there are no differences between both methods; in terms of precision.

F-scan is more restrictive than z-scan because the distance between the sample and the tunable-lens has to be measured with very significant precision. Nevertheless, the distance can be measured with high precision using the nonlinearity of the material; it corresponds to the tunable-lens focal distance at the minimum transmittance.

\section*{Acknowledgements}
E. Rueda thanks Comité para el Desarrollo de la Investigación -CODI-(Universidad de Antioquia - U de A). E. Marulanda thanks Programa Jóvenes Investigadores U de A 2022. E. Rueda thanks the financial support of G8 2020-39036 Universidad de Antioquia, Instituto Tecnológico Metropolitano, Universidad Pontificia Bolivariana, Institución Universitaria Salazar y Herrera, y Ruta N.


 \bibliographystyle{elsarticle-num} 
 \bibliography{refs}

\end{document}